\documentstyle[aps,pra]{revtex}
\tightenlines
\title{Comments on ``On Bohm trajectories in two-particle interference
devices" by L. Marchildon}
\author{Partha Ghose}
\address{S. N. Bose National Centre for Basic Sciences
Block JD, Sector III, Salt Lake, Calcutta 700 098}
\date{\today}
\begin{document}
\maketitle

\begin{abstract}

Marchildon's arguments against my earlier work are refuted.
\end{abstract}

\section{Introduction}

Marchildon \cite{March1} has attempted to show that claims made by
me \cite{Ghose1} and Golshani and Akhavan \cite{GA} that de
Broglie-Bohm (dBB) mechanics and standard quantum theory (SQT) are
incompatible in the case of certain two-particle interference
experiments is wrong. I will restrict my comments strictly to his
criticism of my work, and show that my basic conclusion stands.

\section{Non-ergodic properties of Bohmian motion}

Let me begin by first pointing out that my conclusions are based
on a very well known result in the study of ergodic problems in
mechanics, namely that a decomposable dynamical system is
non-ergodic, and that for such systems the space mean and time
mean are not the same in general. SQT is necessarily always
ergodic whereas Bohmian systems can be non-ergodic, as I will
show. This is the basic reason for their incompatibility.

For the sake of clarity, let me define what is meant by a
decomposable system. Let $(M, \mu, \phi_t)$ be a dynamical system,
where $M$ is a smooth manifold, $\mu$ a measure on $M$ defined by
a continuous positive density, and $\phi_t : M \rightarrow M$ a
one-parameter group of measure preserving diffeomorphisms. If $f$
is a complex-valued function on $M$, its time mean is defined by

\begin{equation}
f^* = {\rm lim}_{N \rightarrow \infty} \frac{1}{N} \sum_{n = 0}^{N
- 1} f (\phi_t^n x),\,\,\,\,\,\,{\rm x\,\,\,\, \epsilon\,\,\,\,
M,\,\, n\,\,\,\, \epsilon\,\,\,\, Z^+}
\end{equation}
and its space mean by

\begin{equation}
{\bar f} = \int_M f (x) d \mu
\end{equation}
A dynamical system is ergodic if for every complex-valued
$\mu$-summable function $f$ the time mean is equal to the space
mean:

\begin{equation}
f^* = {\bar f}
\end{equation}
Now, let $M$ be the disjoint union of two sets $M_1$ and $M_2$ of
positive measure, each of which is invariant under $\phi_t$:

\begin{equation}
\phi_t M_1 = M_1,\,\,\,\,\, \phi_t M_2 = M_2
\end{equation}
Such a system is called a decomposable system. A decomposable
system is not ergodic. Conversely, a non-ergodic system is
decomposable. Finally, a system is ergodic if, and only
if, it is indecomposable. The interested reader is urged to look
at the well known book on ergodic properties of classical
mechanics by Arnold and Avez \cite{Arnold}.

Having cited these well established fundamental results, I will first
prove that dBB is non-ergodic whenever the two-particle
wavefunction $\Psi (x_1, y_1, x_2, y_2, t) = R(x_1, y_1, x_2, y_2,
t){\rm exp}\frac{i}{\hbar}S(x_1, y_1, x_2, y_2, t)$ has the
following symmetries: \vskip 0.1in \noindent(a) exchange or
bosonic symmetry,

\noindent(b) reflection symmetry about an axis (say the $x$ axis
so that $x = {\rm constant}$ is the axis of reflection symmetry),

\noindent(c) translation symmetry of its phase $S$ along that
axis, \vskip 0.1in \noindent and in addition, \vskip 0.1in
\noindent (d) the initial positions of the pair of particles
satisfy the constraint

\begin{equation}
x_1 (0) + x_2(0) = \delta
\label{eq:1}
\end{equation}
where $\delta \approx 0$ is a very small constant, and

\noindent (e) there are no more than a single pair of particles in
the apparatus at any time. \vskip 0.1in

\noindent (A realistic experimental arrangement in which these
conditions can be simulated to a high degree of accuracy will be
discussed in the next section.)

Proof: For any fixed value of $(y_1, y_2)$ $S$ is then a function of
$(x_1 - x_2)$ only, and therefore it follows from the definition
$v_i (x_1, x_2) = {\dot x_i} =
\partial_i S (x_1 - x_2) (i = 1, 2)$ of Bohmian velocities that

\begin{equation}
v_{(1)} (x_1, x_2) + v_{(2)} (x_1, x_2) = {\dot x_1} + {\dot x_2}
= 0 \label{eq:1c}
\end{equation}
and hence

\begin{equation}
x_1 (t) + x_2(t) = x_1 (0) + x_2(0) = {\rm constant} \label{eq:A}
\end{equation}
$\large(x_1 (t)	 + x_2 (t)\large)$ is therefore a constant of motion.

Since the initial positions of the individual pairs in an ensemble
can fluctuate randomly, let us assume that
for every trial lasting a very small time interval around $t_n$,

\begin{equation}
 x_1 (t_n)  + x_2 (t_n) = \delta_n \label{eq:B}
\label{eq:1b}
\end{equation}
where $- d/2 \leq \delta_n \leq d/2$, $d$ being a negligibly small
positive number compared to the (finite) range of $x$ values
accessible to the particles (to be identified with the slit width
in the next section). Then, the symmetry axis $x = \delta_n/2$
will fluctuate from one trial to another but will always remain
bounded within the range $- d/2 \leq x \leq d/2$ around $x = 0$.
Whenever one particle is on the symmetry axis, its partner must
also be on it according to (\ref{eq:B}), and therefore the
$x$-components of their velocities must vanish on the symmetry
axis according to (\ref{eq:1c}). This means the trajectories will
not cross the symmetry axis in any trial, and the Bohmian system
is decomposable and non-ergodic. Q.E.D.

However, since the concept of trajectories does not exist in SQT,
the corresponding system in SQT is indecomposable and ergodic.
Nevertheless, SQT and dBB are by construction compatible in every
conceivable case at the level of space means of observables. This
is achieved in the following way. In SQT, $\vert \Psi (x_1, x_2,
..., x_m, t)\vert^2 \equiv R^2 (x_1, x_2, ..., x_m, t)$ gives the
probability density of finding the $m$ particles distributed in a
particular fashion at a given instant of time if the particles in
the ensemble were to be observed at that instant. In dBB one
introduces a real statistical ensemble of particles with a
probability density $P(x_1(t), x_2(t), ..., x_m(t))$ that is the
same as in SQT. To be precise, dBB postulates that

\begin{equation}
P (x_1(t), x_2(t), ..., x_m(t)) = R^2 (x_1, x_2, ..., x_m, t)
\label{eq:2}
\end{equation}
for all times $t$. This is guaranteed if it holds for the initial
time by virtue of the continuity equation for $P$ that is also
assumed. Then, it can be shown \cite{Holland} that for every
obserbable ${\hat O}$ in SQT (except the momentum ${\hat p}$ which requires special treatment \cite{Bohm}) and the corresponding dynamical
variable $O$ in dBB,

\begin{eqnarray}
\langle {\hat O} \rangle &=& \int ...\int \Pi_{i=1}^{m} dx_i\,\,
\Psi^* {\hat O} \Psi\\\nonumber &=& \int ...\int \Pi_{i=1}^{m}
dx_i\,\, O P (x_1(t), x_2(t), ..., x_m(t)) = {\bar O}
\end{eqnarray}
This demonstrates the equivalence between dBB and SQT for space
means, i.e., averages over the (Gibbs) ensemble of possible states
of the system at a given instant of time. In particular, the
joint detection probability of two-particles defined as a space
mean or ensemble average is given by

\begin{equation}
{\bar P_{1 2}}	= \int_{D_1} \int_{D_2} d x_1 dx_2 R^2 (x_1, x_2,
t)= \int_{D_1} \int_{D_2} d x_1 dx_2 P (x_1(t), x_2(t))
\label{eq:3}
\end{equation}
where $D_1$ and $D_2$ are the supports of the two detector faces.

The situation is different for a time ensemble
when the Bohmian system is decomposable and non-ergodic. In the
two-particle case under consideration, let $x = (x_1, x_2)$. Then

\begin{equation}
\phi_t^n x = \frac{1}{\delta (0)}\ ( x_1(t_n), x_2(t_n))\,\,
\delta ( x_1 (t_n) + x_2 (t_n) - \delta_n )
\end{equation}
and the joint detection probability as a time mean is given by

\begin{equation}
P_{1 2}^{*'} = {\rm lim}_{N \rightarrow \infty} \frac{1}{N}
\sum_{n = 0}^{N - 1} P (\phi_t^n x)\vert_{D_1, D_2}
\end{equation}
which vanishes if the detectors are placed sufficiently
asymmetrically about the $x = 0$ axis (both on the same side of
the symmetry axis, for example). This is in conflict with the SQT
prediction (\ref{eq:3}) for ${\bar P}_{1 2} = P^*_{1 2}$ which
does not vanish for asymmetrically placed detectors. This
concludes the proof of incompatibility.

We will now proceed to see if the conditions (a) through (e) can
be met in an actual experimental set up, and to what degree of
accuracy.

\section{Experimental design criteria}

The above conditions can be very well met by a careful design of
the experimental arrangement. I proposed that a double-slit
arrangement with momentum-correlated particles at the source, so
that single pairs of these particles could pass through two narrow
slits one at a time (condition (e)), would be appropriate.
Therefore, Marchildon's comment at the end of page 3 following
equations (7) and (8) that the "overwhelming majority of pairs are
not simultaneously on the plane of symmetry" does not apply to
this case.

There are, however, two other points of experimental design that
need to be taken care of.

First, the wavefunction very close to the two slits of small width
$d = 2 \delta$ is not plane, and therefore its phase is not a
function of $(x_1 - x_2)$ alone. This means that even if the
particles at the slits satisfy the condition (\ref{eq:B}), when
they enter the Fraunhoffer region with translation invariance,
their positions will no longer satisfy this condition. However,
Marchildon \cite{March2} has shown that with spherical waves
emanating from each point of the slits,

\begin{equation}
\frac{d x_1}{d t} + \frac{d x_2}{d t} = \frac{\hbar k}{m L}(x_1 +
x_2) = \frac{v}{L} (x_1 + x_2)
\end{equation}
where $v = \hbar k/m$ is the velocity of the particles and $L$ is
the distance between the slits and the final screen
or detectors, is a good approximation. Then it follows that

\begin{equation}
x_1 (t) + x_2 (t) = (x_1 (0) + x_2 (0)) e^{v t/L} \leq d e^{v t/L}
\end{equation}
This shows that the exponential factor will grow from unity to the
value $e$ by the time the particles arrive at the detectors. As
long as the Fraunhoffer region sets in much earlier than this, the
particles will enter this region satisfying the condition
(\ref{eq:B}) to an excellent degree of approximation. The
apparatus must be designed to take care of this. This can be done
by a suitable choice of the slit width $d$, the separation $a$
between the slits and the wavelength of the particles.

Second, it is also necessary to ensure that the initial wave
packets do not spread appreciably within the apparatus of length
$L$. Since

\begin{equation}
\frac{\sigma_t}{\sigma_0} = [ 1 + (\frac{\hbar t}{2 m
\sigma_0^2})^2]^{1/2}
\end{equation}
for Gaussian wave packets (equation (17) in \cite{March1}), this
means, for example, that the spreading will be small for electrons
traversing a length $L = 10^2$ cm with a velocity of about
$10^{10}$ cm/sec and initial width $\sigma_0 \approx 2 \times
10^{-4}$ cm. These are neither theoretically impossible nor
unrealistic values.

I must conclude by saying that the actual experiment that is being
planned is with down-converted photons. This is necessitated by
the experimental difficulty of producing a pair of momentum
correlated electrons or neutrons. The experimental design is being
done carefully to take all the above precautions into account. In
fact, one advantage of using photon wavepackets is that they do
not spread. The Bohmian trajectories for this case have been
calculated and can be seen in Ghose {\it et al} \cite{Ghose2}.

\section{Acknowledgement}
I am grateful to Anilesh Mohari for helpful discussions on ergodic
systems and to DST, Govt of India for a research grant that enabled
this work to be undertaken.


\end{document}